\documentclass[prb,reprint,nofootinbib]{revtex4-2} 

\usepackage{amsmath}  
\usepackage{amsfonts} 
\usepackage{graphicx} 

\usepackage[]{hyperref}
\hypersetup{colorlinks=true,linkcolor=blue,citecolor=blue,urlcolor=blue,pdfpagemode=UseNone}

\usepackage[dvipsnames]{xcolor}

\begin{document}

\title{Experimental Visualization of a Fermi Gas}

\author{Inna M. Vishik}
	 \email[Corresponding author: ]{ivishik@ucdavis.edu}
	\affiliation{Department of Physics and Astronomy, University of California, Davis, Davis, California $95616$, USA}
    \affiliation{Materials Sciences Division, Lawrence Berkeley National Lab, Berkeley, CA 94720, USA}

\begin{abstract}
Angle-resolved photoemission spectroscopy (ARPES) can measure electrons' energy vs momentum relations in solids via photoelectric effect. This pedagogical paper experimental results from this technique as a tool for visualizing the Fermi gas model in quantum statistical mechanics. This model is an important starting point for understanding more complex electronic behaviors in real metals.
\end{abstract}

\date{\today}

\maketitle


\section{Introduction}
Angle-resolved photoemission spectroscopy (ARPES) measures the behavior of valence electrons in crystalline solids.  In recent decades it has emerged as a powerful characterization technique in solid state physics, and it is routinely used to understand emergent materials' behaviors via the measured energy vs momentum relations of their electrons. The foundations of the technique\textemdash its operating principle as well as initial understanding of what the data represent \textemdash stem directly from undergraduate concepts, and in turn, the experimental results can help make some undergraduate concepts more tangible.

This paper connects topics from standard undergraduate-level courses in quantum mechanics and statistical mechanics (photoelectric effect, free electron gas) to experimentally measured spectra in real materials. Although ARPES is discussed in some modern graduate textbooks, it is rarely introduced in the standard undergraduate curriculum despite its present-day ubiquity and connection to multiple foundational concepts.

This paper is organized as follows.  First, we review the photoelectric effect with an emphasis on results relevant for interpreting ARPES data. Then, the Fermi gas model of electrons in a metal is presented starting from `particle-in-a-box.'  The results of this calculation are compared to ARPES data on a metal.  Further details of the experimental technique is included for interested readers, and we close with some discussion about the significance of these measurements in the modern research landscape.

\section{Photoelectric effect}
The photoelectric effect, originally explained by Albert Einstein, is normally taught in modern physics courses to motivate the quantization of light energy \cite{Einstein_1905,Millikan_1916}. In the photoelectric effect, light impinges on a material, which the formulas below assume to be a metal.  Electrons can be ejected (photoemitted) only if the photon energy exceeds the material's work function; that is, $hf\geq\Phi$, where $h$ is Planck's constant, $f$ is the frequency of the light, and $\Phi$ is the work function.  The photoelectric effect has two major applications in scientific research: photomultiplier tubes, which are the basis of many ultra-sensitive photon detection schemes, and photoemission spectroscopy, which is the topic of this paper.  

In photoemission, total energy and momentum of the photon and electron are conserved.  This implies that electrons ejected by the photoelectric effect still carry information about their state inside their original material. We will assume in this paper that the material is a crystalline solid, defined by a regular periodic arrangement of atoms, since this will yield a well-defined relationship between energy and momentum of electrons.  Inside the solid, electrons have a binding energy, $E_B$, which relates to the kinetic energy that they have after photoemission ($E_\text{{kin}}$) via a statement of energy conservation:


\begin{equation}
    E_\text{{kin}}=hf-\Phi-|E_B|.
    \label{energy_cons}
\end{equation}

 When considering momentum conservation, the momentum of the photons is ignored, since ultraviolet (UV) photons have relatively small momentum ($p=hf/c$) compared to relevant momentum scales in the experiment. (The same is not true for $x$-ray photoemission \cite{Fadley_2012}, with $hf\gtrsim$ 1000 eV, but this paper concentrates on UV experiments.) In a UV photoemission experiment (typically, $hf \approx$ 20-200 eV), momentum ``conservation" is expressed by equating the momentum of electrons inside the material before photoemission to the momentum of a free electron outside the material after photoemission. The momentum parallel to the sample surface that electrons have inside a material ($p_{||}$ or $\hbar k_{||}$), as measured by photoemission, is given by:
 
\begin{equation}
p_{||}=\hbar k_{||}=\sqrt{2mE_\text{{kin}}}\cdot\sin\vartheta,
\label{momentum_cons}
\end{equation}

\noindent where $m$ is the free electron mass and $\vartheta$ is the emission angle relative to the surface normal.  Between Eqns. \ref{energy_cons} and \ref{momentum_cons}, we have a relationship between binding energy and momentum \textemdash a dispersion relation for electrons inside the solid.  ARPES measures the relationship between energy ($E_B$) and momentum ($k_{||}$) that electrons have inside crystalline solids, via measurements of $E_\text{{kin}}$ and $\vartheta$ of photoemitted electrons.  Electrons in a three-dimensional crystalline solid also have momentum perpendicular to the sample surface ($\hbar k_\perp$), which can be inferred experimentally via more complex procedures \cite{Damascelli_2004}.

\section{Particle-in-a-box and free electron gas}
The photoelectric effect can measure dispersion relations of electrons inside a material, and our starting point for understanding this aspect of electrons' behavior in crystalline solids is the  `particle-in-a-box' (or infinite potential well) model introduced in a quantum mechanics  or modern physics classes.  For simplicity, this model is first reviewed with fixed boundary conditions typically used in these initial courses.  

When a particle is confined to a region of space with length $L$, the continuum of free electron solutions to the Schr\"{o}dinger equation gets pruned to a series of quantized solutions. The standing wave solutions for the infinite potential well with fixed boundary conditions correspond to placing a positive integer ($n$) number of half-wavelength of a harmonic wave inside the length of the well. 
We can express the wavelengths ($\lambda_n$) of these waves inversely in terms of their wavenumber $k_n=2\pi/\lambda_n=n\pi/L$.  It is preferable to express the wavelength this way because $\hbar k$ is the eigenvalue of the momentum operator, and this momentum is related to the one described in Eqn. \ref{momentum_cons}.
The energy solutions to a 1D particle-in-a-box are:
\begin{equation}
E_n=\frac{\hbar^2k^2}{2m}=\frac{\hbar^2n^2\pi^2}{2mL^2}
\label{energy_eigen}
\end{equation}
This energy is all kinetic because the potential energy term is zero inside the `box' where the particles are confined, and the first expression  of the energy can be viewed simply as a classical kinetic energy $E_\text{{kin}}=p^2/2m$, replacing momentum $p$ with quantum mechanical version $\hbar k$. 

In three dimensions, the Schr\"{o}dinger equation is separable, yielding the following energy eigenvalues:
\begin{equation}
E=\frac{\hbar^2}{2m}(k_x^2 +k_y^2+k_z^2)
\label{Evsk}
\end{equation}
where $k_x, k_y, k_z$ are the allowable quantized wavenumbers in the $x$, $y$, and $z$ directions, with the same quantization condition as in 1D.  Distinct ($k_x, k_y, k_z$) coordinates define each distinct solution (wavefunction), but different wavefunctions can correspond to the same energy (degeneracy).

In the context of this paper, the particle-in-a-box model is used to establish the form of the energy eigenvalues used in the Fermi gas model (also called the Sommerfeld model), which is the starting point for understanding the behavior of many electrons in a metal.  It is assumed that the aforementioned single-particle solutions remain unchanged when applied to many particles. Coulomb repulsion between electrons is ignored, as is Coulomb attraction between free electrons and the positively charged ions. Electrons' only interaction is via Pauli exclusion. Amazingly, some metals actually behave in a manner close to this model, which sets the foundation for understanding electrons' energy vs momentum relations in nearly all materials. Now, $L$ becomes the macroscopic dimension of the metal, with implications for spacing between energy levels. When $L$ has approximately atomic length scales (e.g. $<1~ nm$), the spacing between lower energy levels for a confined electron is on the order of $1$ eV. Meanwhile, if $L$ is taken to be macroscopic (e.g. $1 ~cm$), energy spacing diminishes to $\approx 10^{-14}$ eV, beyond the detection limit of electron spectroscopies. Thus, the energy levels in this thermodynamic limit effectively behave like a continuum.

Before going further, we return to the topic of boundary conditions.  Because we ultimately want to make connections between the Fermi gas model and real metals, we need to modify the boundary conditions from those chosen to meet students' assumed familiarity to those more appropriate for our experimental context.  In solid state physics it is advantageous to use periodic, not fixed, boundary conditions because \cite{ashcroft1976solid} 1) they allow surface effects to be ignored; 2) they admit traveling, not standing, wave solutions, which are needed for charge and heat transport; and 3) they admit the sorts of wavefunctions (``Bloch waves") that are consistent with a periodic lattice in a crystalline solid. The latter two are beyond the scope of this paper, but because we want to connect to experimental observations in metals, they are stated for completeness. 

The plane wave solutions from periodic boundary conditions that we use here are: $\psi_k(x, y, z)=e^{i(k_xx+k_yy+k_zz)}$ where $k_{x,y,z}$ can take on positive or negative values \footnote{The more complete form of the wavefunction (written in 1D for brevity) is $\psi_k(x)=A e^{i k x} + B e^{-i k x}$ where $A$ and $B$ are constants. The form of the wavefunction used here is standard in solid state physics textbooks \cite{ashcroft1976solid} and sets $B=0$. The form with two complex exponentials yields the same quantized values of $k$, once periodic boundary conditions for both $\psi_k(x)$ and its derivate $\partial \psi_k (x)/\partial x$ are considered.}.

In the Fermi gas model, we take all of the relevant electrons in the metal and place them into the available states, implicitly at zero temperature, following these rules:
\begin{itemize}

    \item Each state, defined by a different spatial wavefunction (distinct $k_x, k_y, k_z$), can hold two electrons of different spin states. Electrons are fermions with spin quantum number taking on two possible values, $-1/2$  or $+1/2$. This Pauli exclusion is the only mutual repulsion for electrons in this model.
	\item Lowest energy states are filled first, then higher energies are filled sequentially.  In three dimensions, according to Eqn. \ref{Evsk}, $(k_x,k_y,k_z)$ coordinates with the same energy form a spherical shell in $k$-space.  Thus, filling in order of ascending energy corresponds to filling up a sphere one shell at a time.
    \item When we have used up all the electrons, we reach the Fermi energy ($E_F$).  $E_F$ is the highest energy filled (occupied) state at zero temperature.  This corresponds to a value (magnitude) of $k$ called the ‘Fermi momentum’, $k_F$, where $E_F=(\hbar^2 k_F^2)/2m$.
\end{itemize}

The quantization of $k$ will be used to rewrite the expression for $E_F$. Presently we assume a cubic box of length $L$ on each side and periodic boundary conditions.  Along the $x$ direction we have:
\begin{align}
    &\psi(x=0)=\psi(x=L)\\
    &e^{i k_x \cdot 0}=1=e^{i k_x \cdot L}\\
    &k_x=0, \pm2\pi/L, \pm4\pi/L,...
\end{align}
And same for $k_y, k_z$

\begin{figure}[h]
    \centering
    \includegraphics[width=1\linewidth]{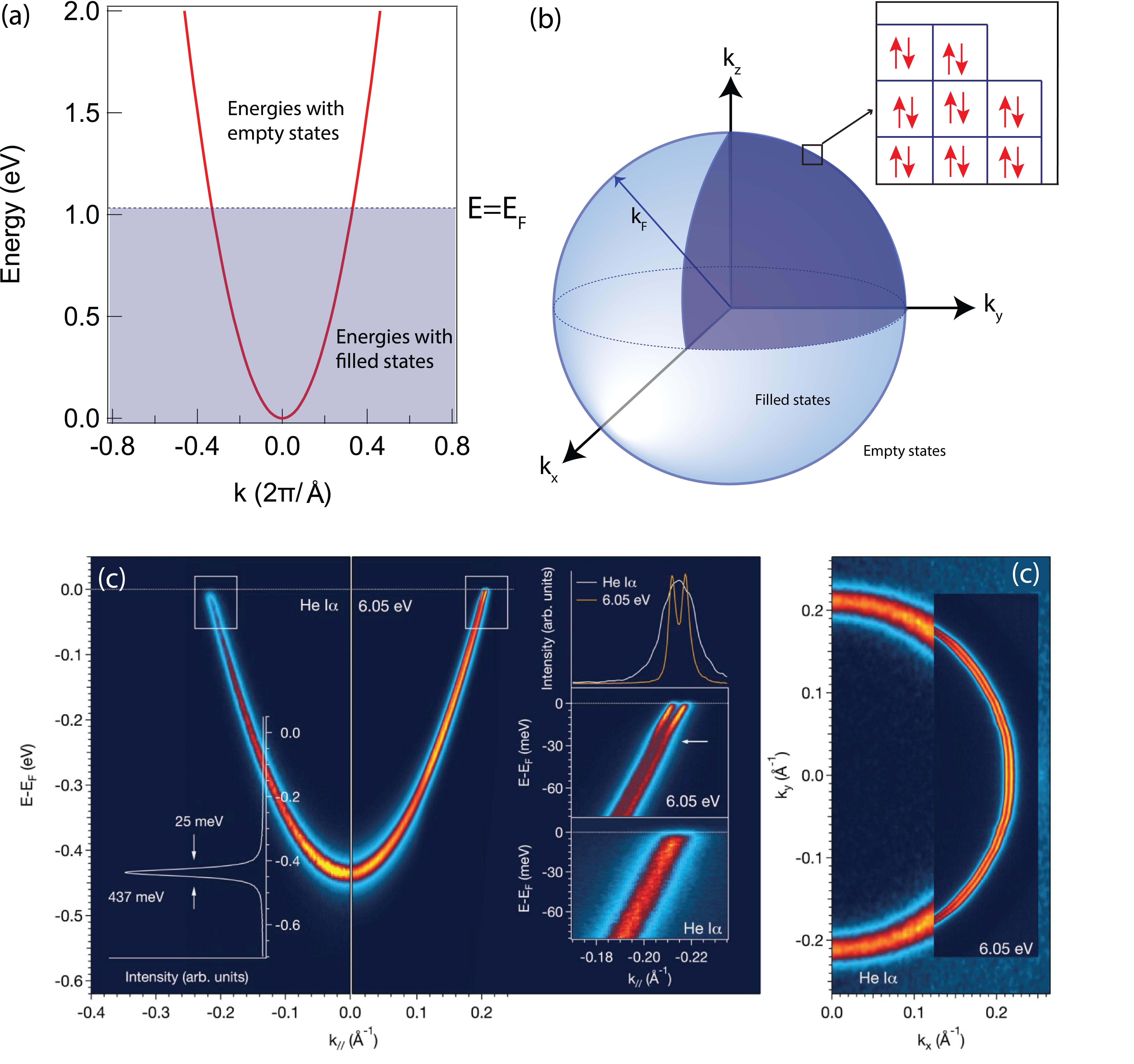}
    \caption{(color online) Fermi gas, theory and experiment. (a) Energy vs $k$ for free electron gas, with occupied energy levels below $E=E_F$.  The Fermi level is chosen to roughly correspond to one electron per $64 \AA^3$, which is a reasonable electron density for a 3D metal.  (b) Sketch of Fermi gas at $T=0$ in 3D $k$-space showing filled states within sphere of radius $k_F$. Inset shows momentum states filled by electrons of opposite spin. (c) ARPES spectrum on 2D electron system (surface state of copper), Ref. \onlinecite{Tamai:2013} showing Energy vs momentum. He-I$\alpha$ and 6.05 eV refer to two different lightsources, a helium plasma lamp and a UV laser, respectively, used to measure the same system sequentially.  Lower left inset is a vertical slice at $k_{||}=0$ indicating the bottom of the parabola at $E_B=$ 437 meV.  Insets on right show horizontal slices at $E_F$ for the two lightsources, and zoomed-in details of two boxes in main panel. (d) Half of the constant energy map at $E_F$ for the same 2D metal. Discontinuity of linewidth of circle originates from different resolution of the two lightsources. Measurement temperature is 6K. Panels (c)-(d) reproduced with permission from Phys. Rev. B 87, 075113 (2013). Copyright 2013, American Physical Society.}
    \label{fig:Cu111}
\end{figure}

Now we use the constraint of the fixed number of electrons, N.  For a real metal, this number should be the number of valence electrons.  For example, for sodium, this number will be one per atom. Our sphere in $k$-space has volume $\frac{4}{3}\pi k_F^3$, and this sphere is divided into voxels of volume $(2 \pi /L)^3$. Each voxel can hold two electrons of opposite spin (Fig. \ref{fig:Cu111}(b)). Thus, twice the number of Voxels yields the number of electrons: 
\begin{equation}
N=2\frac{(4/3)\pi k_F^3}{(2 \pi /L)^3}.
\end{equation}
Next, we solve for $k_F$ and use this to write an equation for $E_F$: 
\begin{equation}
E_F=\frac{\hbar^2}{2m}(\frac{3\pi^2 N}{V})^{2/3}.
\end{equation}
This equation also uses that volume $V=L^3$. Usually, the assumption that the box was cubic can be relaxed \cite{ashcroft1976solid}, so that we are left with an expression in terms of electron density, $N/V$, the electron mass, and physical constants.  For typical electron densities in 3D, $E_F\approx 1-5 eV$, which is very large compared to room-temperature thermal energy, given by Boltzmann constant multiplied by temperature, $k_B T$. Thus, this ground state treatment remains qualitatively accurate at experimentally-accessible temperatures. The expressions for $E_F$ will be different in 2 and 1 dimensions, but the derivations are equivalent. 

The standard derivation above is an extension of a particle-in-a-box model to a macroscopic ``box" that holds many electrons, with assumptions of minimal interactions.  One of the most exciting aspects of experimental physics is to be able to visualize these textbook concepts in real physical systems.  This is what is shown in Fig. \ref{fig:Cu111} which presents ARPES spectra on a simple 2D metal from the recent published literature, in this case, a copper surface state \cite{Tamai:2013}. ARPES spectra are typically presented as color plots, where the highest-intensity regions (orange in Fig. \ref{fig:Cu111}) correspond to energy and momentum values where an electron is most likely to be extracted.  In Eqn. \ref{Evsk} and Fig. \ref{fig:Cu111}(a), we see that the free particle Fermi gas model has a quadratic relationship between energy and momentum; in Fig. \ref{fig:Cu111}(c), we see in ARPES data that the most intense (orange) part of the energy vs momentum spectrum forms a parabolic shape. This parabola appears continuous, owing to the vanishingly small energy between quantized energy levels, and it stops abruptly at energy $E_F$, reflecting that this is the highest energy occupied state. Another aspect of the Fermi gas model is that shells of constant energy form a sphere in 3D or a circle in 2D $k$-space (Fig. \ref{fig:Cu111}(b)).  This is seen in ARPES data in Fig. \ref{fig:Cu111}(d), where a 2D momentum map at $E_F$ show the high intensity region forming a circle, and radius of this circle defines $k_F$. $k_\perp$ is absent from this discussion because the system being measured is two dimensional, existing only on the surface of a copper crystal. In solid state physics, this shape formed by the highest-energy occupied states in the momentum map is called a `Fermi surface.' Fermi surfaces are significant because they are the origin of the charge and heat transport properties of a metal.  The $k$-space construction of the Fermi gas model is sometimes not intuitive for students used to thinking in real space, but such structures can be visualized directly with these experiments based on the photoelectic effect.

Fig. \ref{fig:Cu111} also illustrate how improvements in instrument resolution can produce sharper spectra, which in turn reveal additional physics.  The right-hand insets of Fig. \ref{fig:Cu111} (c) reveals two features instead of one, when measured with the 6.05 laser which offers better energy and momentum resolution, and this splitting originates from coupling between the spin and orbital motion of electrons \cite{LaShell_1996}.  Additionally, there is a small bend, marked by the arrow, which is interpreted in terms of electrons interacting with atomic vibrations (phonons).  This demonstrates how more advanced physics concepts about electrons' behavior in materials can build on top of the basic Fermi gas model, with signatures of both being observable in a single experiment.

\section{ARPES experimental setup}
\begin{figure}[h!]
    \centering
    \includegraphics[width=1\linewidth]{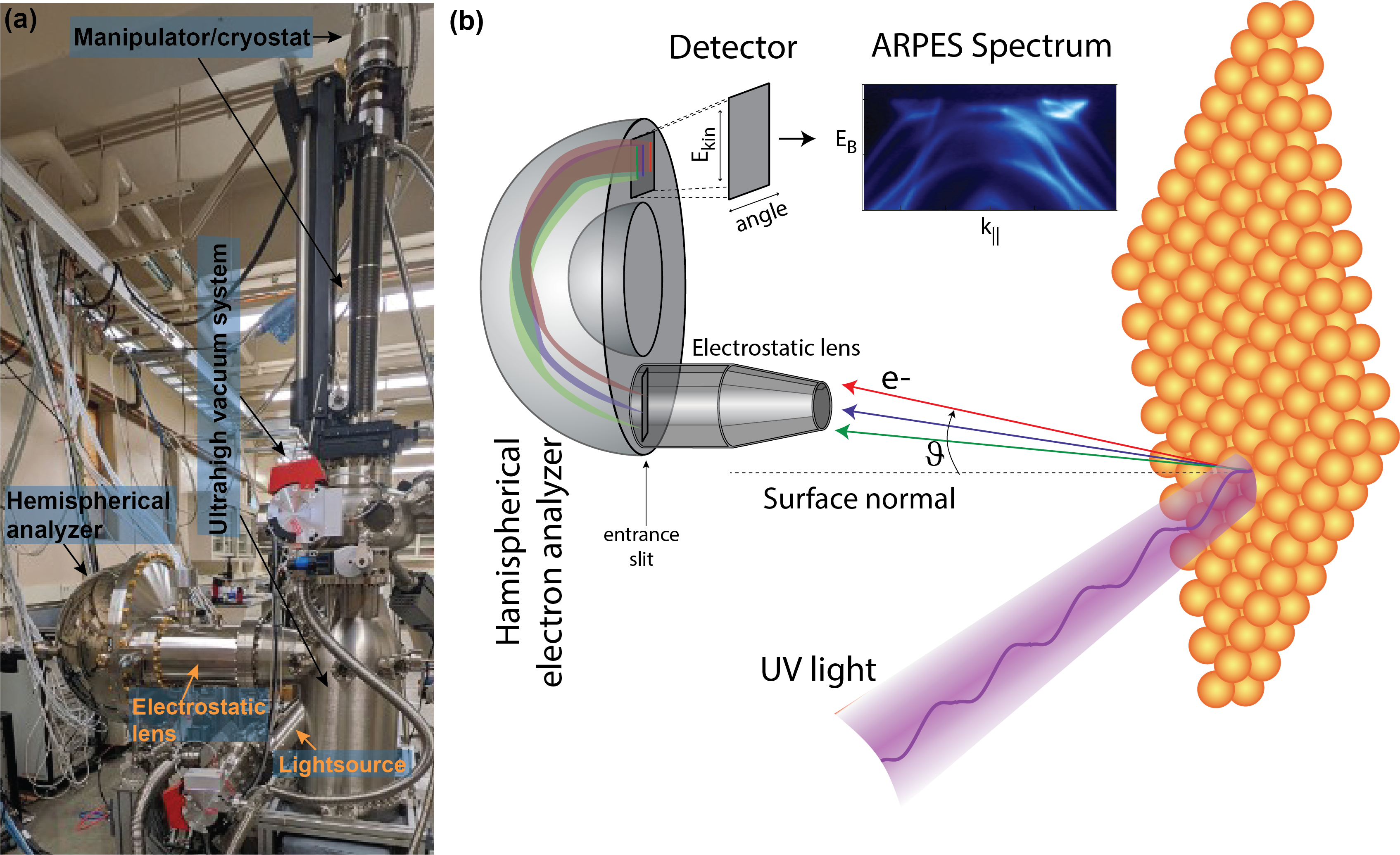}
    \caption{ARPES experiment. (a) Example experimental lab-based ARPES system showing Helium-lamp lightsource, UHV system, the analyzer system consisting of electrostatic lens column and hemispherical analyzer, and sample manipulator/cryostat. (b) Schematic of ARPES experiment showing crystalline sample and emission angle $\vartheta$ referenced to surface normal. Different emission angles are imaged onto different positions on entrance slit before entering hemispherical analyzer.  ARPES spectrum uses data from Ref. \onlinecite{Rossi_WTe2_2020}}
    \label{fig:setup}
\end{figure}

Now that the fundamentals of the quantities being measured have been introduced, we turn to a brief discussion of the components of ARPES experiments \cite{Zhang_2022,Wang_2024}. An example lab-based system is shown in Fig. \ref{fig:setup} (a). The necessary components include: 1) high power UV lightsources 2) ultrahigh vacuum (UHV) system 3) the electron analyzer which enables simultaneous measurement of electrons' kinetic energy and emission angle, and 4) a manipulator/goniometer and cryostat to control sample position/orientation and temperature \cite{Sobota:RMP_2021}. A schematic of the experimental geometry and key experimental components is shown in Fig. \ref{fig:setup}(b).

An intense lightsource is critical to measuring photoemitted electrons with sufficient signal-to-noise ratio. For UV photoemission, the primary light sources are lasers (e.g. 6.05 eV panels in Fig. \ref{fig:Cu111}(c)-(d)), plasma lamps (e.g. He-I$\alpha$ panels in Fig. \ref{fig:Cu111}(c)-(d) and Fig. \ref{fig:setup}(a)), and synchrotrons; the light source defines $hf$ in Eqn. \ref{energy_cons}. The first two types of lightsources are housed in labs or small shared facilities, and the latter is a type of large user facility accessed via a proposal process. Synchrotron-based experiments are currently the modal experimental route \cite{Magdy_2025} because the broadly tunable photon energy is an important tuning knob that allows, for example, to measure $k_{\perp}$ \cite {Damascelli_2004} or highlight features originating from different orbitals \cite{Moser_2017}.

ARPES experiments have stringent requirements for the pressure inside the experimental chamber, typically operating at $<10^{-10}$ Torr, more than 12 orders of magnitude smaller than atmospheric pressure.  This requirement exists because UV ARPES is extremely surface sensitive \cite{Seah_1979}. For electrons to maintain their momentum information after photoemission, they must exit the material without interacting with any atoms or other electrons therein; they can only do this if they originate close to the surface (roughly within $1 ~nm$). Thus, a fresh surface is prepared prior to each measurement \cite{Zhang_2022}, and even under UHV, molecules from the chamber will adsorb (`stick') onto the specimen, eventually obscuring the fresh surface  (typically within 24-48 hours).

The electron analyzer is the most sophisticated part of the experiment, and the most common type is hemispherical \cite{Wang_2024}.  It consists of  a series of electrostatic lenses that use electric fields to map emission angles of the electrons emitted from the sample \cite{M_rtensson_1994} onto specific spatial positions on an entrance slit. Afterwards, the hemispherical part of the analyzer, which consists of two concentric hemispheres held at different electric potentials (a parallel plate capacitor), guides electrons to a detector, with electrons having different kinetic energy assuming a trajectory with different radius. Because this is a precision measurement of low energy electrons, magnetic shielding is important.  Electrons are imaged on a two-dimensional detector with axes corresponding to $E_\text{{kin}}$ and angle along the entrance slit.  The the detector image can look almost exactly like published spectra (e.g. Fig. \ref{fig:Cu111}(c)), except the axes need to be adjusted using Eqns. \ref{energy_cons} and \ref{momentum_cons} (Fig. \ref{fig:setup}(b)).

Finally, the sample being measured is grounded, to replenish photoemitted electrons, and sits on a manipulator/cryostat --- a movable stage which allows to vary the temperature and move the sample to the proper measurement position/angle.

\section{Conclusions and Outlook}
This paper has discussed how the photoelectric effect finds application in a popular contemporary research tool, ARPES, and how the standard quantum statistical mechanical model of a free electron gas can be visualized with this research tool in some simple metals.   

The power of the ARPES technique is that it can measure electrons' energy vs momentum (dispersion) relations, $E(k_x,k_y,k_z)$. Without interactions other than Pauli exclusion, the dispersion relation is set by the kinetic energy of free electrons, which is observed experimentally in some real metals (Fig. \ref{fig:Cu111}). When interactions (e.g. with the crystalline lattice, with other electrons, or with excitations stemming from atomic vibrations or magnetism) are considered, they will change the energy eigenvalues and dispersion relations, which is also directly measurable with the same technique. Canonical examples include electrons that behave like ``light" with linear relationships between energy and momentum \cite{Sprinkle:GrapheneARPES2009}, electrons that behave as if their mass is much larger than the free electron mass \cite{Chen_2017_CeCoIn5}, and carriers that behave as if they had positive instead of negative charge \cite{Wu_2015_WTe2}. 

Dispersion relations are significant for understanding a material's electronic properties because 1) they readily connect to quantities such as electrons' group velocity and effective mass, which influence how they respond to electromagnetic fields or temperature and 2) because novel phenomena in materials often manifest in changes to dispersion relations. 

ARPES is used to reveal phenomena at the cutting edge of solid state physics, but the first steps to interpreting the spectra come from undergraduate modern and statistical physics courses: the photoelectric effect and particle-in-a-box extended to a free electron gas.  In the current research landscape, undergraduate students get involved with ARPES experiments in myriad ways including: synthesizing new materials to be studied by the technique, performing the experiments themselves, participating in constructing and maintaining lab-based ARPES experiments, and analyzing the information-rich data sets, many of which are now publicly available.








{\textbf{Acknowledgments}  This work was supported by the National Science Foundation (US) Division of Materials Research Award. No. 2428464. }

{\textbf{Conflict Disclosure Statement:} The authors have no conflicts to disclose.}

\bibliography{References}

\end{document}


\setlength{\intextsep}{5pt}
\title{\Large
Supplementary Materials}

\author{ Inna M. Vishik $^{1,2}$*}

\affil[1]{\small Department of Physics and Astronomy, University of California, Davis, CA 95616, USA}
\affil[2]{Materials Sciences Division, Lawrence Berkeley National Lab, Berkeley, California 94720, USA}

\date{}
\maketitle
\vspace{-6ex}

\setcounter{secnumdepth}{0}

These supplementary materials includes content that was cut from the published paper for brevity, but may be useful to student researchers.

\section{Fermi-Dirac distribution}
The careful reader will notice that the parabola in Fig. 1(c) stops suddenly at $E_F$ or that the derivation in the previous section was strictly for temperature $T=0$, which is not accessible in the laboratory. At elevated temperature, the Fermi energy generalizes to the chemical potential, $\mu$, which is a thermodynamical concept that reflects the quantity that is equal once two systems that can exchange particles come to equilibrium and stop exchanging particles; it has units of energy.  But notice that the y-axis of Fig 1(c) is in terms of $E_F$ even though temperature not zero. This is a colloquialism in the field, using $E_F$ at non-zero temperature. It is justified by the fact that usually $\mu\approx E_F$ for lab-accessible temperatures. 

The final piece of describing the basic attributes of an ARPES spectrum is the Fermi-Dirac distribution.  The derivation is briefly reviewed below, following the content of a statistical mechanics course.  For a distribution of quantum particles where the number of particles can vary, the partition function is given by the following expressions, where the second part includes the fact that electrons are fermions, so the only possible occupation of a state is 1 or 0
\begin{align}
&\mathcal{Z}=\Sigma_n e^{-n(\epsilon-\mu)/k_BT}\\
&\mathcal{Z}_{fermion}=e^{-1(\epsilon-\mu)/k_BT}+e^{-0(\epsilon-\mu)/k_BT}\\
&\mathcal{Z}_{fermion}=e^{-(\epsilon-\mu)/k_BT}+1
\end{align}
Where $n$ is the number of particles in each state, $\epsilon$ is the energy of each state, and $k_B$ is the Boltzmann constant. The probability of a state of energy $\epsilon$ to be occupied by $n$ electrons is given by: 
\begin{equation}
P(n)=\frac{1}{\mathcal{Z}}e^{-n(\epsilon-\mu)/k_BT}
\end{equation}
The average occupation ($\bar{n}$) is given by:
\begin{equation}
\bar{n}=\Sigma_n nP(n)=\frac{1}{1+e^{(\epsilon-\mu)/k_BT}}\equiv f(\epsilon)
\label{FD}
\end{equation}

\begin{figure}[h]
    \centering
   \includegraphics[width=\linewidth]{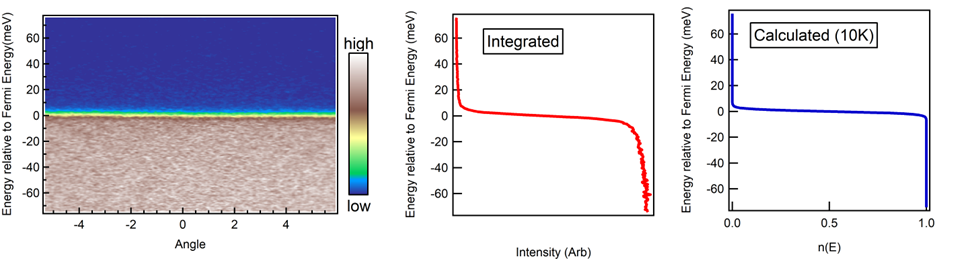}
   \caption{Fermi-Dirac distribution. (a) ARPES measurement on polycrystalline gold at 10K (b) spectrum in (a) integrated over all vertical slices (c) calculated Fermi-Dirac distribution at 10K}
    \label{fig:FD_10K}
\end{figure}

$f(\epsilon)$ is the Fermi-Dirac distribution, also called the Fermi function, and it is plotted in Fig. \ref{fig:FD_10K}(c) at $T=10 K$.  It is shaped like a rounded step function where the rounding subtends a wider energy range at higher temperature. Physically, the Fermi function dictates that states at much lower energy than $\mu$ are completely filled and states at much higher energy are completely empty.  It is only within the vicinity of $\mu$ where occupation is not entirely boring, and it is this narrow energy range which is responsible for all low energy excitations that allow electrons to conduct electricity and heat.

The Fermi function multiplies every ARPES spectrum, such that only the energy vs momentum relations of `occupied' states are visible. If one measures a disordered polycrystalline metal (Fig. \ref{fig:FD_10K}(a)-(b)), the result reflects a Fermi function.  These measurements are often performed for calibration, to experimentally determine $E_F$ (or more precisely, $\mu$).

\section{Electrons beyond free electron gas}
Dispersion relations, also called `electronic band structure' give detailed information about electrons' motion in a crystalline solid. Two quantities relevant to motion are velocity and mass, and the procedure for extracting them from dispersion relations is discussed below.  Importantly, these quantities can be extracted and are meaningful \textit{even if dispersion relations are not free-electron like}.    

Extracting velocity: the momentum operator acting on a free-electron wavefunction yields a momentum eigenvalue $\mathbf{p}=\hbar \mathbf{k}$.  Dividing by mass gives
\begin{equation}
\mathbf{v}=\hbar \mathbf{k}/m
\label{group_v_free}
\end{equation}
We now generalize this result to scenarios outside free electrons.  For a general wavepacket, the group velocity (in 1D) is given by $v_g=\frac{\partial \omega}{\partial k}$ were $\omega$ is an angular frequency where $E=\hbar \omega$.  Thus, for energy vs momentum dispersion relations, $v_g=\frac{1}{\hbar}\frac{\partial E}{\partial k}$, which generalizes in 2D and 3D to 
\begin{equation}
    \mathbf{v_g}=\frac{1}{\hbar}\nabla_k E(k)
    \label{group_velocity}
\end{equation}
For the free-electron dispersion, we can see that this produces the same result as dimensional analysis in Eqn. \ref{group_v_free}.  However, the result in Eqn. \ref{group_velocity} is generally true even if free-electrons are not a good starting point.  For electrons at $E_F$, the extracted velocity is called the Fermi velocity, $v_F$.  Because electrons very close to $E_F$ are the only ones capable of low-energy excitations into unoccupied states, $v_F$ is crucial for quantifying how electrons transport charge and heat.

Because of interactions, electrons inside crystalline solids often behave as if they have a mass different from the free electron mass.  This experimental effective mass ($m^*$) can also be extracted from dispersion relations. We will do this in two ways, which are equivalent for free electron dispersions, the first method is again dimensional analysis:
\begin{equation}
m^* v(k) = \hbar k \rightarrow m^* =\hbar k/v (k)
\label{mass_from_k}
\end{equation}
The second method is from the curvature of the dispersion relation:
\begin{equation}
m^*=\frac{1/\hbar^2}{\frac{\partial^2E}{\partial k^2}}
\label{mass_from_curature}
\end{equation}
For a parabolic dispersion, $m^*$ will be a single value for all values of $k$, while other cases will yield momentum-dependent mass; if this occurs, the value at $E_F$ is the one relevant for corresponding with transport and thermodynamic measurements.

\begin{figure}[h]
    \centering
    \includegraphics[width=\linewidth]{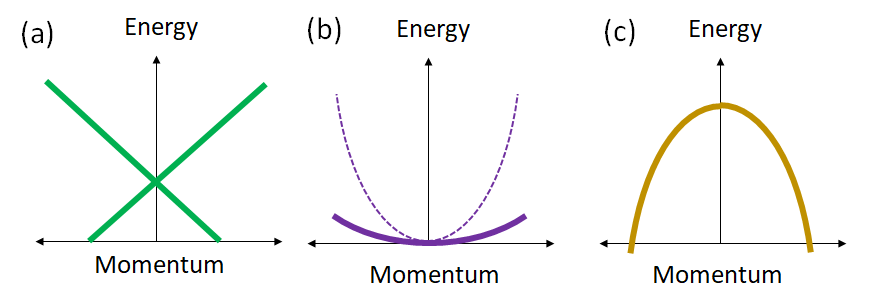}
    \caption{Examples dispersion relations, schematics. (a) linear (Dirac-like) dispersion.(b)'heavy' dispersion. (c) 'hole-like' dispersion. }
    \label{fig:examples}
\end{figure}

Real materials have a variety of dispersion relations, and a few examples are discussed below.  A popular motif in recent decades has been materials whose electrons move like light, with linear relationship between energy and momentum.  With light, the proportionality constant is the speed of light, but of course electrons in a solid cannot travel as fast as light, so instead, the proportionality constant is the Fermi velocity. Graphene is the canonical compound showing linear (or Dirac-like) dispersion, but similar features are found in many compounds. Oppositely, in some materials electrons can behave as if they have masses much larger than the free electron mass.  Because of the mass term in the denominator of Eqn. (4), such electrons have dispersion relations that look `flat.' Example canonical materials with flat bands include heavy fermion compounds, Moire materials, and Kagome metals.  In some materials, the electrons' dispersions look like a parabola flipped upside down.  This corresponds to charge carriers behaving as if they had opposite (positive) charge from electrons, and indeed, these 'holes' respond to electromagnetic fields opposite to electrons. 

These three examples are just a few of the possibilities of how every material is its own universe with different speeds of light, different electron mass, and sometimes with antimatter dominating over matter. 
